\documentclass[a4paper]{jpconf}
\usepackage[square,sort&compress,numbers]{natbib}
\usepackage{graphicx}
\usepackage{amsmath,amssymb,amsfonts}
\usepackage{algorithmic}
\usepackage{textcomp}
\usepackage{xcolor}

\begin{document}

\title{Classical Monte Carlo algorithm for simulation of a pseudospin model for cuprates}

\author{V A Ulitko$^1$, Yu D Panov$^1$, A S Moskvin$^1$}

\address{$^1$Institute of Natural Sciences and Mathematics, Ural Federal University, 19 Mira str., Ekaterinburg , Russia }

\ead{vasiliy.ulitko@urfu.ru}

\begin{abstract}
A classical Monte Carlo algorithm based on the quasi-classical approximation is applied to the pseudospin Hamiltonian of the model cuprate.
The model takes into account both local and non-local correlations, Heisenberg spin-exchange interaction, single-particle and correlated two-particle transfer.
We define the state selection rule that gives both the uniform distribution of states in the phase space and the doped charge conservation. 
The simulation results show a qualitative agreement of a phase diagrams with the experimental ones. 
\end{abstract}

\section{Introduction}
The phase diagram of doped HTSC cuprates is the subject of active experimental~\cite{Bozovic2016,Bozovic2019} and theoretical research, despite the huge amount of work on this topic to date. 
A striking feature of the phase diagram of HTSC cuprates is the competition and coexistence of antiferromagnetic, superconducting, and charge orderings~\cite{Fradkin2012}, manifested in pseudogap phase, strange metal phase, a variety of static and dynamic fluctuations. 
The studies are complicated by the presence of heterogeneity due to dopants or non-isovalent substitution, as well as to the internal electronic tendency to heterogeneity~\cite{Moskvin2019}. 
Phase separation may be the cause of simultaneous detection of the preformed pairs and BEC superconductivity in cuprates~\cite{Bozovic2019}, a number of experimental observations of the typical Fermi liquid behavior, at least in overdoped cuprates. 
For models describing such complex multiphase states, the calculation of phase diagrams within the exact schemes is obstructed due to the absence of one leading parameter, and therefore, to obtain physically reliable results it is natural to use straightforward techniques, such as the mean field approximation and the classical Monte Carlo method. 

Previously, we developed a minimal model of the HTSC cuprates~\cite{Moskvin2011,Moskvin2013}, where the CuO$_2$ planes are considered as lattices of centers, which are the main element of the crystal and electronic structure of cuprates.
In this model, on-site Hilbert space is formed by three effective valence states of the CuO$_4$ cluster: [CuO$_4$]$^{7-}$, [CuO$_4$]$^{6-}$, and [CuO$_4$]$^{5-}$. 
The necessity to consider these valence states of CuO$_4$ center on an equal basis is related to the strong relaxation effects of the electron lattice in cuprates~\cite{Mallett2013,Moskvin2019-2}. 
The valence states of CuO$_4$ center have different spin states: $s=1/2$ for the [CuO$_4$]$^{6-}$ center and $s=0$ for the [CuO$_4$]$^{7-}$ and [CuO$_4$]$^{5-}$, respectively, 
and different symmetry of the orbital states: $B_{1g}$ for the ground states of the [CuO$_4$]$^{6-}$ center, $A_{1g}$ for the [CuO$_4$]$^{7-}$ center, and the Zhang-Rice $A_{1g}$ or more complicated low-lying non-Zhang-Rice states for the [CuO$_4$]$^{5-}$ center. 
For these many-electron states with strong $p{-}d$ covalence and strong intra-center correlations, 
electrons cannot be described within conventional (quasi)particle approach 
that addresses the [CuO$_4$]$^{7-,6-,5-}$ centers within the on-site hole representation 
$|n\rangle$, $n = 0, 1, 2$, respectively. 
We make use of a real space on-site $S=1$ pseudospin formalism to describe the charge triplets instead of conventional quasiparticle $k$-momentum description. 
The pseudospin approach is used for the strongly correlated electron systems~\cite{Castellani1979,Rice1981} and for the superconductivity~\cite{Low1994} of cuprates for a long time. 
In our model, the effective pseudospin Hamiltonian takes into account both local and nonlocal correlations, single and two-particle transport, as well as Heisenberg spin-exchange interaction. 
Earlier, we investigated a simplified static version of the spin-pseudospin model, for which phase diagrams of the ground state and at a finite temperature were constructed, both analytically, in the mean field approximation~\cite{Panov2019}, 
and as a result of Monte Carlo simulations~\cite{Yasinskaya2020}. 
The use of pseudospin formalism provides opportunities for numerical modeling using the well-developed classical Monte Carlo (MC) method, the construction of phase diagrams and the study of the features of the thermodynamic properties of the system. 
A similar effective $S=1$ spin-charge model for cuprates and its MC implementation 
were considered in papers~\cite{Cannas2019,Frantz2021}.

We organize the article as follows. 
In Section 2, we present the pseudospin formalism and the effective spin-pseudospin Hamiltonian of the model and introduce quasi-classical approximation. 
In Section 3, we formulate the state selection algorithm and explore the features of the probability distribution.
The results of classical MC simulations of our model and their discussion are presented in Section 4.

\section{Model}

We develop a pseudospin model of cuprates~\cite{Moskvin2011,Moskvin2013} where the CuO$_2$ planes are considered as lattices of CuO$_4$ clusters, which are the main element of the crystal and electronic structure of cuprates. 
The on-site Hilbert space is formed by 4 states. 
The effective valence states of the cluster, 
[CuO$_4$]$^{7-}$, [CuO$_4$]$^{6-}$, and [CuO$_4$]$^{5-}$, 
have different spin states: formally one-hole [CuO$_4$]$^{6-}$ center is the $s = 1/2$ doublet, 
while the [CuO$_4$]$^{7-}$ and [CuO$_4$]$^{5-}$ centers are the spin singlets.
As a result, the basis $\left| SM; s\mu \right\rangle$ on a given site is the quartet of states
$\big\{ \left| 11; 00 \right\rangle $, $\left| 10; \frac12 \frac12 \right\rangle $, 
$\left| 10; \frac12, {-}\frac12 \right\rangle$, $\left| 1,-1; 00 \right\rangle \big\}$. 

The effective pseudospin Hamiltonian of the model cuprate
\begin{equation}
	\mathcal{H} = \mathcal{H}_{pot} 
	+ \mathcal{H}_{kin}^{(1)} + \mathcal{H}_{kin}^{(2)} 
	+ \mathcal{H}_{ex} 
	\label{eq:Ham0}
\end{equation}
takes into account both local and nonlocal charge correlations 
\begin{equation}
	\mathcal{H}_{pot} = \sum_i \left( \Delta S_{zi}^2 - \mu S_{zi} \right) 
	+ V \sum_{\left\langle ij\right\rangle} S_{zi} S_{zj} , 
	\label{eq:Hpot}
\end{equation}
the three types of the correlated single-particle transport 
\begin{equation}
	\mathcal{H}_{kin}^{(1)}
	\;=\;
	- \sum_{\left\langle ij\right\rangle \nu} 
		\left[ t_p  P_{i{+}}^{\nu} P_{j{-}}^{\nu} 
		+  t_n  N_{i{+}}^{\nu} N_{j{-}}^{\nu}  
		+  \frac{t_{pn}}{2} 
		\left(P_{i{+}}^{\nu} N_{j{-}}^{\nu} + N_{i{+}}^{\nu} P_{j{-}}^{\nu}\right) 
		+ h.c. \right] , 
	\label{eq:Hkin1}
\end{equation}
the two-particle transport 
\begin{equation}
	\mathcal{H}_{kin}^{(2)}
	=	- t_b \sum_{\left\langle ij\right\rangle} 
	\big(  S_{i{+}}^2 S_{j{-}}^2 + S_{j{+}}^2 S_{i{-}}^2  \big) , 
	\label{eq:Hkin2}
\end{equation}
and finally, the antiferromagnetic Heisenberg spin-exchange interaction for the CuO$_4^{6-}$ centers, 
\begin{equation}
	\mathcal{H}_{ex} = Js^2 \sum_{\langle ij \rangle} \boldsymbol{\sigma}_i \boldsymbol{\sigma}_j , 
	\label{eq:Hex}
\end{equation}
where $\boldsymbol{\sigma}=P_0 \, \mathbf{s}/s$ operators take into account the on-site spin density $P_0 = 1-S_z^2$, 
and $\mathbf{s}$ is the spin $s=1/2$ operator. 
The pseudospin operator $S_z$ in \eqref{eq:Hpot} 
gives the value of charge counted from ''parent'' [CuO$_4$]$^{6-}$ state on a given site, 
so the term with chemical potential $\mu$ allows to account for the charge density constraint, 
$nN = \left\langle \sum_i S_{zi} \right\rangle = const$. 
Operators $P_{{+}}^{\nu}$ in \eqref{eq:Hkin1} create holes with the spin projection $\nu$ 
and change the states $\left| 00;\frac{1}{2},{-}\nu \right\rangle$ 
into the states $\left|11;00\right\rangle$. 
Likewise, operators $N_{{+}}^{\nu}$ also create holes with the spin projection $\nu$, but 
they transform the states $\left| 1,{-}1;00 \right\rangle$ 
into $\left| 00;\mbox{$\frac{1}{2}$} \nu \right\rangle$. 
Operators $S_{{+}}^2$ in \eqref{eq:Hkin2} creates the singlet hole pairs 
on the [CuO$_4$]$^{7-}$ centers, 
and, obviously, the following relations for the one-hole and two-hole creation operators are fulfilled: 
$S_{+}^2 = P_{+}^{\nu} N_{+}^{-\nu}$. 
The explicit form of matrices for operators in equations (\ref{eq:Hpot}--\ref{eq:Hex}) 
in the basis of states $\left| SM; s\mu \right\rangle$ is given in Appendix.

\section{Critical temperatures of the ''pure'' phases in the mean fied approximation}

In the mean-field approximation (MFA) in \cite{Panov2019pmm}, the equations of critical temperatures of ''pure'' ordered phases with only one non-zero order parameter were found. 
We introduce two sublattices $A$ and $B$ forming a staggered order on the square lattice. 
For some operator $\hat{C}_{i}$ its average depends on the sublattice index, $\big\langle \hat{C}_{i} \big\rangle = C_{\alpha}$, $i \in \alpha=A,B$. 

For the charge-ordered (CO) phase with the order parameter $L=\left(S_{z,A}-S_{z,B}\right)/2$ the equation for the critical temperature $T_{CO}$ has the form
\begin{equation}
	T = \frac{4 V \left( 1 - n^2 \right) \phi(n,T)}{1 + \phi(n,T)} , 
	\label{eq:TCO}
\end{equation}
where 
\begin{equation}
	\phi(n,T) = \sqrt{ \left( 1 - n^2 \right) e^{-2\Delta / T} + n^2 } . 
\end{equation}
This equation generalizes the equation for the critical temperature of charge ordering in the hard-core bosons model \cite{Micnas1990}.
The concentration dependence of $T_{CO}$ 
is $4 V \left( 1 - n^2 \right)$ at $\Delta / V \to-\infty$,
and at $\Delta / V \geq 2$ it tends to the limiting value $T_{CO}=4V|n|\left(1-|n|\right)$

For the antiferromagnetic (AFM) phase, the order parameter is defined as $\boldsymbol{l} = \big( \boldsymbol{\sigma}_{A} - \boldsymbol{\sigma}_{B} ) / 2$. 
The critical temperature $T_{AFM}$ can be found from the equation 
\begin{equation}
	T = \frac{4 J s^{2} \left( 1 - n^2 \right) }{1 + \phi(n,T)} . 
	\label{eq:TAFM}
\end{equation}
At $\Delta\geq0$, the concentration dependence of $T_{AFM}$ has a maximum at $n=0$, changing from $T_{AFM}=2Js^2\left(1-n^2\right)$ at $\Delta=0$ to $T_{AFM}=4Js^2\left(1-\left|n\right|\right)$ at $\Delta\rightarrow+\infty$. 

By analogy with the model of local bosons \cite{Micnas1990}, the phase with a nonzero average $\big\langle \hat{S}_{+}^2 \big\rangle$  can be called a bose superfluid (BS). 
Equation for the critical temperature $T_{BS}$
\begin{equation}
	T = 4 t_{b} n \left[ \ln \frac{\left(1 + n\right)\left(\phi(n,T) + n\right)}{\left(1 - n\right)\left(\phi(n,T) - n\right)} \right]^{-1} 
	\label{eq:TBS}
\end{equation}
generalizes the known result \cite{Micnas1990} and leads to an expression for $T_{BS}$ in the model of local bosons at $\Delta\to -\infty$. 

By analogy, for phases with non-zero order parameters 
$\big\langle \hat{P}_{m}^{+} \big\rangle$ and $\big\langle \hat{N}_{m}^{+} \big\rangle$  we can find in the case $t_{pn}=0$ the equations for critical temperatures $T_p$:
\begin{equation}
	T = 2 t_{p} \frac{ \left(1+n\right)\left[1-2n-\phi(n,T)\right] }{ \left[ 1+\phi(n,T) \right] \ln \left( \frac{1 - n}{\phi(n,T) + n} \right) } , 
	\label{eq:Tp}
\end{equation} 
and $T_{n}$:
\begin{equation}
	T = 2 t_{n} \frac{ \left(1-n\right)\left[1+2n-\phi(n,T)\right] }{ \left[ 1+\phi(n,T) \right] \ln \left( \frac{1 + n}{\phi(n,T) - n} \right) } . 
	\label{eq:Tn}
\end{equation} 
In these phases, correlated single-particle transport of hole (P) or electron (N) type is realized.

\section{The energy in quasi-classical approximation}

Using the quasi-classical approximation, we write the on-site wave function as follows 
\begin{equation}
	\left|\Psi\right\rangle
	= c_{1} \left| 11; 00 \right\rangle 
	+ c_{\uparrow} \left| 10; \tfrac12 \tfrac12 \right\rangle 
	+ c_{\downarrow} \left| 10; \tfrac12, {-}\tfrac12 \right\rangle 
	+ c_{{-}1} \left| 1,-1; 00 \right\rangle , 
	\label{eq:psi}
\end{equation}
where the complex coefficients can be written in the following form:
\begin{equation}
	c_{{k}} = r_k \, e^{i\phi_k} , \qquad
	\sum_k r_k^2 = 1 , 
	\label{eq:ck}
\end{equation}
with phases $\phi_k \in [0,2\pi]$, 
and we parametrize magnitudes $r_k$ by angles 
$\theta,\varphi,\psi \in [0,\frac{\pi}{2}]$: 
\begin{eqnarray}
	r_{1} &=& \cos\theta \cos\varphi , \label{eq:r1} \\
	r_{\uparrow} &=&  \sin\theta \cos\psi , \label{eq:r12} \\
	r_{\downarrow} &=&  \sin\theta \sin\psi , \label{eq:rm12} \\
	r_{{-}1} &=&  \cos\theta \sin\varphi . \label{eq:rm1} 
\end{eqnarray}
The average values for all operators in the Hamiltonian \eqref{eq:Ham0} are given in Appendix.  

The energy for a model \eqref{eq:Ham0} 
in the quasi-classical approximation
\begin{equation}
	E = \Big\langle \prod_i \Psi_i \Big| \, \mathcal{H} \, \Big| \prod_i \Psi_i \Big\rangle
\end{equation}
have the following form:
\begin{multline}
	E = \sum_i \left(\Delta - \mu \cos 2\varphi_i \right) \cos^2 \theta_i 
	+ V \sum_{\left\langle ij \right\rangle} 
	\cos^2 \theta_i  \cos 2\varphi_i  \cos^2 \theta_j  \cos 2\varphi_j 
	-{} \\
	 - \frac{t_p}{2}  \sum_{\left\langle ij \right\rangle}  
	 \sin 2\theta_i  \cos \varphi_i  \sin 2\theta_j  \cos \varphi_j 
		\Big(  \cos \psi_i  \cos \psi_j 
		\cos \left(\phi_{1i}-\phi_{\uparrow i}-\phi_{1j}+\phi_{\uparrow j}\right) 
	+{} \\
	+ \sin \psi_i  \sin \psi_j  
		\cos \left(\phi_{1i}-\phi_{\downarrow i}-\phi_{1j}+\phi_{\downarrow j}\right)  \Big) 
	-{} \\
	 - \frac{t_n}{2}  \sum_{\left\langle ij \right\rangle}  
	\sin 2\theta_i  \sin \varphi_i  \sin 2\theta_j  \sin \varphi_j  
		\Big(  \cos \psi_i  \cos \psi_j  
		\cos \left( \phi_{{-}1i}-\phi_{\uparrow i}-\phi_{{-}1j}+\phi_{\uparrow j}\right) 
	+{} \\
		+ \sin \psi_i  \sin \psi_j  
		\cos \left(\phi_{{-}1i}-\phi_{\downarrow i}-\phi_{{-}1j}+\phi_{\downarrow j}\right)  \Big) 
	-{} \\
	- \frac{t_{pn}}{4}  \sum_{\left\langle ij \right\rangle}  
	\sin 2\theta_i  \sin 2\theta_j  
		\Big[  
			\cos \varphi_i  \sin \varphi_j  
			\Big(  \sin \psi_i  \cos \psi_j  
				\cos \left( \phi_{1i}-\phi_{\downarrow i}+\phi_{{-1}j}-\phi_{\uparrow j} \right) 
	+{} \\
				+ \cos \psi_i  \sin \psi_j  
				\cos \left( \phi_{1i}-\phi_{\uparrow i}+\phi_{{-}1j}-\phi_{\downarrow j} \right) 
			\Big) 
	+{} \\
			+ \sin \varphi_i  \cos \varphi_j  
			\Big(  
				\sin \psi_i  \cos \psi_j  
				\cos \left( \phi_{{-}1i}-\phi_{\downarrow i}+\phi_{1j}-\phi_{\uparrow j} \right)
	+{} \\
				+ \cos \psi_i  \sin \psi_j  
				\cos \left( \phi_{{-}1i}-\phi_{\uparrow i}+\phi_{1j}-\phi_{\downarrow j} \right) 
			\Big)
		\Big]
	-{} \\
	- \frac{t_b}{2}  \sum_{\left\langle ij \right\rangle}  
		\cos^2 \theta_i  \sin 2\varphi_i  \cos^2 \theta_j  \sin 2\varphi_j 
			\cos \left(\phi_{{-}1i}-\phi_{1i}-\phi_{{-}1j}+\phi_{1j} \right)
	+{} \\
	+ J s^2  \sum_{\left\langle ij \right\rangle}  
		\sin^2 \theta_i  \sin^2 \theta_j  
			\Big( 
			\sin 2\psi_i  \sin 2\psi_j  
			\cos \left( \phi_{\uparrow i}-\phi_{\downarrow i}-\phi_{\uparrow j}+\phi_{\downarrow j} \right) 
			+ \cos 2\psi_i  \cos 2\psi_j 
			\Big) . 
	\label{eq:Efull}
\end{multline}

\section{State selection algorithm}

The magnitudes of coefficients $r_k$ in Eq.~\eqref{eq:ck} correspond to points in the octant of the 4-dimensional unit sphere. 
In the Metropolis algorithm, randomly generated states should form a uniform distribution in the phase space. 
For the parametrization (\ref{eq:r1}--\ref{eq:rm1}), the solid angle element is 
$d\Omega = \cos\theta \sin\theta \, d\theta \, d\varphi \, d\psi$, 
thus, the state selection algorithm should consist of generation of uniformly distributed phases $\phi_k \in [0,2\pi]$, 
uniformly distributed angle variables $\varphi, \psi \in [0,\pi/2]$, 
and uniformly distributed value $ m = \cos^2 \theta \in [0,1]$, where $\theta \in [0,\pi/2]$. 
In this case, the MC simulation of model \eqref{eq:Efull} involves using the chemical potential $\mu$ as external fixed parameter and the subsequent recalculation of the results in the variables charge density, $n$, and temperature, $T$.

\begin{figure}[t]
\centerline{\includegraphics[width=\textwidth]{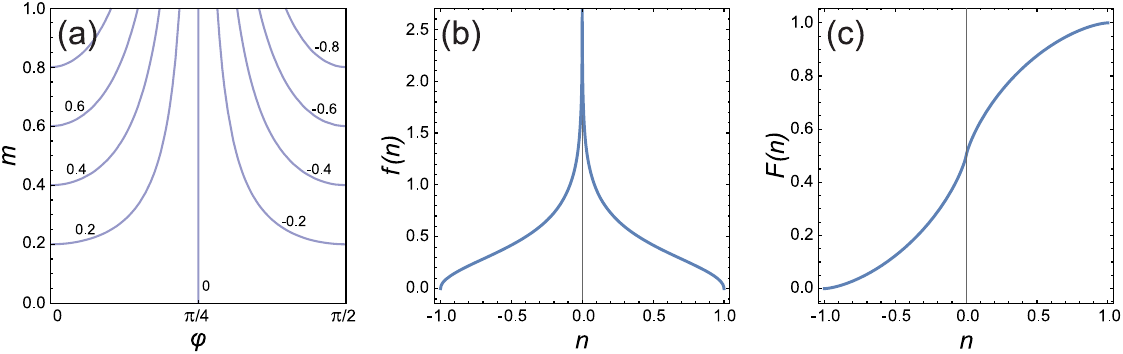}}
\caption{(a) The constant value lines for the on-site charge density $n$ 
defined by Eq.~\eqref{eq:neq}; 
(b) the probability density function $f(n)$; 
(c) the probability distribution function $F(n)$.}
\label{fig:fig1}
\end{figure}

To study the features of the parametrization (\ref{eq:ck}--\ref{eq:rm1}) we can find the on-site charge density distribution which is generated by the state selection algorithm formulated above. 
For the on-site charge density, we obtain the following expression in terms of uniformly distributed variables $\varphi$ and $m$: 
\begin{equation}
	n = r_{1}^2 - r_{{-}1}^2 = m \cos 2\varphi . 
	\label{eq:neq}
\end{equation}
The domains $D(n)$ where $m \cos 2\varphi < n$ are shown in Fig.~\ref{fig:fig1}(a). 
Integrating over domain $D(n)$, we find the on-site charge distribution function $F(n)$ 
\begin{equation}
	F(n) = \frac{2}{\pi} \int_{D(n)} dm \, d\varphi
	= -\frac{1}{\pi} \arccos n + \frac{n}{\pi} \ln \frac{1+\sqrt{1-n^2}}{|n|} , 
	\label{eq:Fn}
\end{equation}
and the corresponding probability density function $f(n)$ 
\begin{equation}
	f(n) = \frac{1}{\pi} \ln \frac{1+\sqrt{1-n^2}}{|n|} . 
	\label{eq:fn}
\end{equation}
These functions are shown in Fig~\ref{fig:fig1}(b,c). 
As a specific feature of the parametrization (\ref{eq:ck}--\ref{eq:rm1}),  
the probability density $f(n)$ has a logarithmic singularity at $n=0$.

One of the phase states in model \eqref{eq:Ham0} is the charge ordering. 
In this case, the function $n(\mu)$ has a typical step-like feature, 
when a small change in $\mu$ causes a large jump in $n$, from $n_1$ to $n_2$, so, 
taking into account the statistical nature of the Monte Carlo method, 
it is difficult to obtain trustworthy simulation results for the range $(n_1,n_2)$. 
Further, we will consider an algorithm where the lattice state changes simultaneously on a pair of sites, but the total charge of the pair is conserved.
This ensures the conservation of the total charge of the system during the simulation and allows us to study in detail the phase states of the system for all $n$.

\begin{figure}[t]
\centerline{\includegraphics[width=\textwidth]{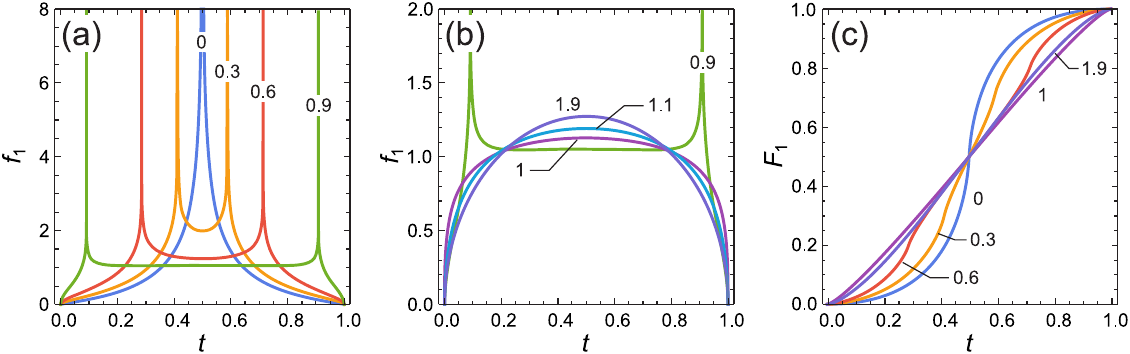}}
	\caption{The normalized probability density function $f_1(t;2n)$ 
	for values of the pair charge 
	(a) $2|n| = 0.0$, $0.3$, $0.6$, $0.9$; 
	(b) $2|n| = 0.9$, $1.0$, $1.1$, $1.9$; 
	(c) the cumulative distribution function $F_1(t;2n)$ 
	for $2|n| = 0.0$, $0.3$, $0.6$, $1.0$, $1.9$. 
	}
	\label{fig:fig2}
\end{figure}

If the states of a pair of sites 1 and 2 generated independently, 
the probability density to have the charge of the pair $2n = n_1 + n_2$
for a given charge $n_1$ at the site 1 is 
\begin{equation}
	f_1(n_1;2n) = \frac{f(n_1)f(2n-n_1)}{\Phi(2n)}
\end{equation}
where
\begin{equation}
	\Phi(2n) = \int_{n_{1,min}}^{n_{1,max}} f(x) f(2n-x) \, dx , 
\end{equation}
and the function $f(n)$ is defined by Eq.~\eqref{eq:fn}. 
The minimal and maximal values of $n_1$ at given $2n$ are 
\begin{equation}
	n_{1,min}(2n) = -1 + n + |n| , \qquad
	n_{1,max}(2n) = 1 + n - |n| .
\end{equation}
The cumulative distribution function $F_1(n_1;2n)$ of the charge $n_1$ at the site 1 for the fixed pair charge $2n$ has the following form:
\begin{equation}
	F_1(n_1;2n) = \int_{n_{1,min}}^{n_{1}} f_1(x;2n) \, dx . 
	\label{eq:F1}
\end{equation}
The normalized probability density function 
$f_1(t;2n) = \Delta n_1 f_1 \left(\Delta n_1 t + n_{1,min}; 2n\right)$, 
where $\Delta n_1 = n_{1,max} - n_{1,min}$, 
and cumulative distribution function $F_1(t;2n)$ are shown in Fig.~\ref{fig:fig2}. 
The probability density function $f_1$ has logarithmic singularities if $2|n|<1$ 
as shown in Fig.~\ref{fig:fig2}(a), 
and the corresponding distribution function $F_1$ has vertical tangents at these points. 
If $1\leq2|n|<2$, the probability density function has no singularities, 
so the distribution function only slightly deviates from the case of uniform distribution.

\begin{figure}[t]
\centerline{\includegraphics[width=\textwidth]{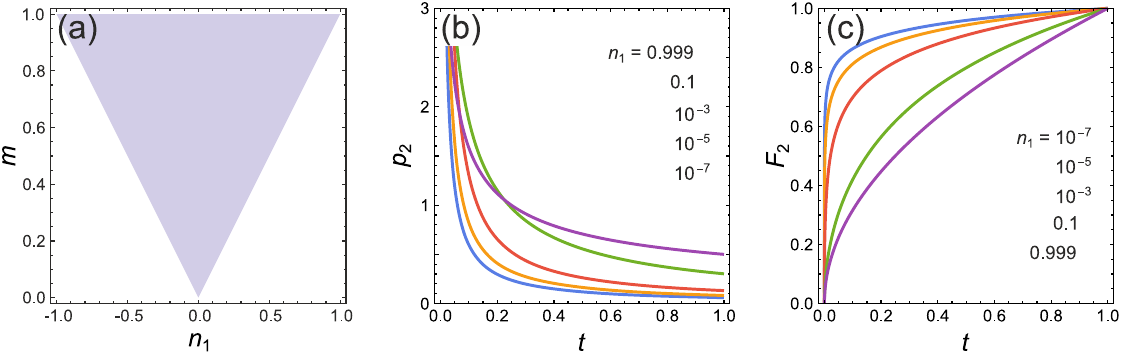}}
	\caption{ 
	(a) The shaded area is the domain of functions in variables $(n_1 , m )$; 
	(b) the conditional density function $p_2$ for given values of $n_1$; 
	(c) the conditional distribution function $F_2$ for given values of $n_1$. 
	}
	\label{fig:fig3}
\end{figure}

The uniform distribution in a phase space entails the constant probability density function 
$f(\varphi , m) = 2/\pi$ in the domain 
$0\leq m \leq 1$, $0 \leq \varphi \leq \tfrac{\pi}{2}$ shown in Fig.~\ref{fig:fig1}(a). 
Since one of the new variables must be $n_1$, we choose them as $\left(n_1 , m \right)$. 
The domain in variables $(\varphi , m)$ is mapped onto the domain in variables $(n_1 , m )$ 
shown in~Fig.\ref{fig:fig3}(a). 
The new density function $p(n_1,m)$ is defined from equations
\begin{equation}
	p(n_1,m) \, dn_1 \, dm  
	= \frac{2}{\pi} \left| \frac{\partial\varphi}{\partial n_1} \right| \, dn_1 \, dm 
	= \frac{ dn_1 \, dm }{\pi \sqrt{m^2 - n_1^2}} . 
\end{equation} 
This allows us to find the conditional density function,  
\begin{equation}
	p_2(m|n_1) = \frac{1}{\pi f(n_1)  \sqrt{m^2 - n_1^2}} , 
\end{equation}
and the conditional distribution function:
\begin{equation}
	F_2(m|n_1) = \frac{ \ln \left(m+\sqrt{m^2-n_1^2}\right) - \ln |n_1| }{ \ln \left(1+\sqrt{1-n_1^2}\right) - \ln |n_1| } , \quad
	|n_1| \leq m \leq 1 . 
	\label{eq:F2}
\end{equation}
Fig.\ref{fig:fig3}(b,c) show the normalized conditional density function 
$p_2(t|n_1) = a\,p_2(at+|n_1||n_1)$, $a=(1-|n_1|)$, 
and corresponding conditional distribution function $F_2(t|n_1)$ 
for some values of $n_1$. 
The most significant variations of these functions take place in the region of small values of the parameter $n_1$, therefore, values decreasing on a logarithmic scale are considered. 
For the state selection algorithm, it is necessary so solve an equation $F_2(m|n_1) = \gamma$ at given $n_1$, so small values of $n_1$ can potentially lead to large inaccuracies. 
Fortunately, the explicit solution of equation $F_2(m|n_1) = \gamma$ can be written:
\begin{equation}
	m = \frac{1}{2} \left[ |n_1|^{1-\gamma} \left(1+\sqrt{1-n_1^2}\right)^{\gamma} 
	+ |n_1|^{1+\gamma} \left(1+\sqrt{1-n_1^2}\right)^{-\gamma} \right] . 
	\label{eq:mSol}
\end{equation}

The state selection algorithm for the quasi-classical Monte Carlo simulation of the model 
\eqref{eq:Ham0} that conserves the total charge consists of the following steps:

\begin{enumerate}
	\item calculation of the total charge $2n = n_{1,0} + n_{2,0}$ for the randomly selected pair of sites $1$ and $2$; 
	
	\item calculation of the value $n_1$ from equation 
	$F_1(n_1;2n) = \gamma$, 
	where $\gamma \in [0,1]$ is the uniformly distributed random value, 
	and the function $F_1(n_1;2n)$ is defined by Eq.~\eqref{eq:F1}; 
	
	\item calculation of the value $n_2 = 2n - n_1$; 
	
	\item calculation of values $m_i$, $i=1,2$, from equations $F_2(m_i|n_i) = \gamma_i$, 
	where $\gamma \in [0,1]$ is the uniformly distributed random value, 
	the function $F_2(m|n)$ is defined by Eq.~\eqref{eq:F2}, and
	the explicit solution is given by~Eq.\eqref{eq:mSol};
	
	\item calculation of $ \varphi_i $ $i=1,2$, from equations $\cos(2\varphi_i) = n_i/m_i$;
	
	\item calculation of $\theta_i$, $i=1,2$, from equations $\cos^2\theta_i = m_i$; 
	
	\item generation of uniformly distributed random values $\phi_k^{(i)} \in [0,2\pi]$, $i=1,2$, 
$k = {+}1,{-}1,\uparrow,\downarrow$, and $\psi_i \in [0,\tfrac{\pi}{2}]$, $i=1,2$. 
\end{enumerate}
This allows us to find new states on the selected pair of sites using Eq.~\eqref{eq:psi}.

\section{Results}

In MC simulation, we calculated the structure factors
\begin{equation}
	F_{\mathbf{q}}(A,B) = \frac{1}{N^2} \sum_{lm} e^{i\mathbf{q}\,(\mathbf{r}_l - \mathbf{r}_m)}
	\left\langle A_{l} B_{m} \right\rangle
	,
\end{equation}
where $A_{l}$ and $B_{m}$ are the on-site operators and the summation is performed over all sites of the square lattice.
To determine the type of ordering, we monitored the following structure factors:
\begin{itemize}
	\item 
	$F_{(\pi,\pi)}(\boldsymbol{\sigma},\boldsymbol{\sigma})$ for antiferromagnetic (AFM) order,
	\item 
	$F_{(\pi,\pi)}(S_z,S_z)$ for the charge order (CO),
	\item 
	$F_{(0,0)}(S_{{+}}^2,S_{{-}}^2)$ for the bose-superluid order (BS),
	\item 
	$F_{(0,0)}(P^{+},P)$ for the ``metal'' P-type phase (P).
\end{itemize}

\begin{figure}[t]
	\centerline{\includegraphics[width=0.75\textwidth]{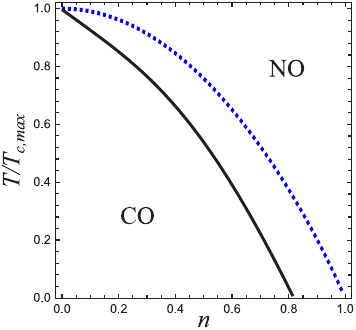}}
	\caption{Critical temperature of CO ordering. 
		The nonzero model parameters are $\Delta=0.1$, $V=0.25$.
		The dotted line shows the MFA value obtained from Eq.~\eqref{eq:TCO}. 
		The solid line corresponds to results of MC simulation. 
	}
	\label{fig:co-phasediag-scaled}
\end{figure}

\begin{figure}[t]
	\centerline{\includegraphics[width=0.75\textwidth]{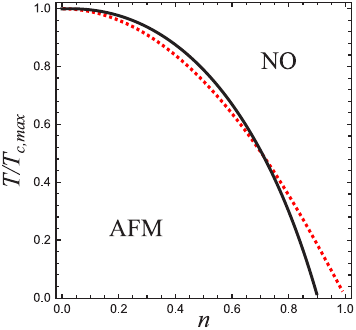}}
	\caption{Critical temperature of AFM ordering.
		The nonzero model parameters are $\Delta=0.1$, $J=1$.
		The dotted line shows the MFA value obtained from Eq.~\eqref{eq:TAFM}. 
		The solid line corresponds to results of MC simulation.
	}
	\label{fig:afm-phasediag-scaled}
\end{figure}

\begin{figure}[t]
	\centerline{\includegraphics[width=0.75\textwidth]{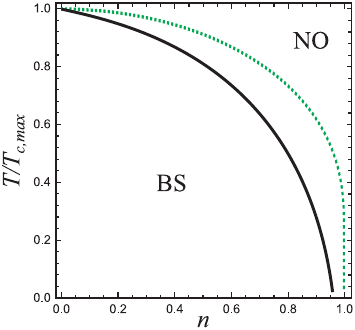}}
	\caption{Critical temperature of BS ordering.
		The nonzero model parameters are $\Delta=0.1$, $t_{b}=1$.
		The dotted line shows the MFA value obtained from Eq.~\eqref{eq:TBS}. 
		The solid line corresponds to results of MC simulation.
	}
	\label{fig:bs-phasediag-scaled}
\end{figure}

\begin{figure}[t]
	\centerline{\includegraphics[width=0.75\textwidth]{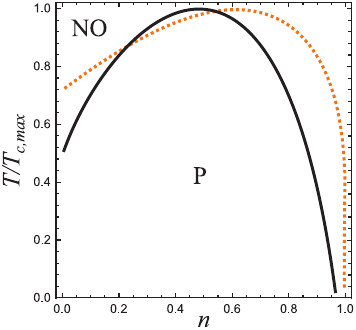}}
	\caption{Critical temperature of the P phase ordering.
		The nonzero model parameters are $\Delta=0.1$, $t_{p}=1$.
		The dotted line shows the MFA value obtained from Eq.~\eqref{eq:Tp}. 
		The solid line corresponds to results of MC simulation.
	}
	\label{fig:fl-phasediag-scaled}
\end{figure}

The results of numerical simulations and comparisons with MFA are shown in Figs.~\ref{fig:co-phasediag-scaled} -- \ref{fig:fl-phasediag-scaled}.
The critical temperature in the MC simulation was determined by reaching the corresponding structural factor of 0.05 of the value at $T\to0$ for a given $n$, 
the region of smaller values of structural factor is designated as the non-ordered (NO) phase.

Taking into account fluctuations in MC simulations within the framework of the quasi-classical approximation used here leads to a significant decrease in critical temperature compared with MFA.
The maximum value of the critical temperature at $\Delta=0.1$ is achieved at $n=0$ for the CO, AFM, and BS phases and at some intermediate value of $n$ for the P phase.
The maximum value ratios for MC and MFA methods, 
$\tau_{c} = T_{c,max}^{MC}/T_{c,max}^{MFA}$, 
are following:
\begin{equation}
\tau_{CO} \simeq 0.27, \quad
\tau_{AFM} \simeq 0.36, \quad
\tau_{BS} \simeq 0.22, \quad
\tau_{P} \simeq 0.14.
\end{equation}

To compare the qualitative behavior of the concentration dependences of the critical temperatures of the ''pure'' phases for the MC and MFA methods, the Figs.~\ref{fig:co-phasediag-scaled} -- \ref{fig:fl-phasediag-scaled} show the values given relative to the maximum.
The results of MC simulation preserves the qualitative form of the concentration dependences of the critical temperatures of the ''pure'' phases, but leads to the appearance of the region of values $n$ in which the ordering does not occur even for $T\to0$.

\section*{Acknowledgments}
The research was supported by the Russian Science Foundation, grant no. 24-21-20147.


\section*{Appendix}

The matrices of pseudospin operators on a given site in the basis 
$\big\{ \left| 11; 00 \right\rangle $, $\left| 10; \frac12 \frac12 \right\rangle $, 
$\left| 10; \frac12, {-}\frac12 \right\rangle$, $\left| 1,-1; 00 \right\rangle \big\}$ 
have the following form:
\begin{equation}
	{
	\setlength\arraycolsep{3pt}
	S_z =
	\begin{pmatrix}
	1 & 0 & 0 & 0 \\[-2pt]
	0 & 0 & 0 & 0 \\[-2pt]
	0 & 0 & 0 & 0 \\[-2pt]
	0 & 0 & 0 & -1 
	\end{pmatrix}
	,\;
	S_z^2 =
	\begin{pmatrix}
	1 & 0 & 0 & 0 \\[-2pt]
	0 & 0 & 0 & 0 \\[-2pt]
	0 & 0 & 0 & 0 \\[-2pt]
	0 & 0 & 0 & 1 
	\end{pmatrix}
	,\;
	S_{+}^2 =
	\begin{pmatrix}
	0 & 0 & 0 & 1 \\[-2pt]
	0 & 0 & 0 & 0 \\[-2pt]
	0 & 0 & 0 & 0 \\[-2pt]
	0 & 0 & 0 & 0 
	\end{pmatrix}
	,\;
	S_{-}^2 =
	\begin{pmatrix}
	0 & 0 & 0 & 0 \\[-2pt]
	0 & 0 & 0 & 0 \\[-2pt]
	0 & 0 & 0 & 0 \\[-2pt]
	1 & 0 & 0 & 0
	\end{pmatrix}
	,
	}
\end{equation}

\begin{equation}
	{
	\setlength\arraycolsep{3pt}
	P_{+}^{\downarrow} = 
	\begin{pmatrix}
	0 & 1 & 0 & 0 \\[-2pt]
	0 & 0 & 0 & 0 \\[-2pt]
	0 & 0 & 0 & 0 \\[-2pt]
	0 & 0 & 0 & 0	
	\end{pmatrix}
	,\;
	P_{-}^{\downarrow} = 
	\begin{pmatrix}
	0 & 0 & 0 & 0 \\[-2pt]
	1 & 0 & 0 & 0 \\[-2pt]
	0 & 0 & 0 & 0 \\[-2pt]
	0 & 0 & 0 & 0 
	\end{pmatrix}
	,\;
	P_{+}^{\uparrow} = 
	\begin{pmatrix}
	0 & 0 & 1 & 0 \\[-2pt]
	0 & 0 & 0 & 0 \\[-2pt]
	0 & 0 & 0 & 0 \\[-2pt]
	0 & 0 & 0 & 0	
	\end{pmatrix}
	,\;
	P_{-}^{\uparrow} = 
	\begin{pmatrix}
	0 & 0 & 0 & 0 \\[-2pt]
	0 & 0 & 0 & 0 \\[-2pt]
	1 & 0 & 0 & 0 \\[-2pt]
	0 & 0 & 0 & 0 
	\end{pmatrix}
	,
	}
\end{equation}

\begin{equation}
	{
	\setlength\arraycolsep{3pt}
	N_{+}^{\uparrow} = 
	\begin{pmatrix}
	0 & 0 & 0 & 0  \\[-2pt]
	0 & 0 & 0 & 1 \\[-2pt]
	0 & 0 & 0 & 0 \\[-2pt]
	0 & 0 & 0 & 0 
	\end{pmatrix}
	,\;
	N_{-}^{\uparrow} = 
	\begin{pmatrix}
	0 & 0 & 0 & 0 \\[-2pt]
	0 & 0 & 0 & 0 \\[-2pt]
	0 & 0 & 0 & 0 \\[-2pt]
	0 & 1 & 0 & 0 
	\end{pmatrix}
	,\;
	N_{+}^{\downarrow} = 
	\begin{pmatrix}
	0 & 0 & 0 & 0  \\[-2pt]
	0 & 0 & 0 & 0 \\[-2pt]
	0 & 0 & 0 & 1 \\[-2pt]
	0 & 0 & 0 & 0 
	\end{pmatrix}
	,\;
	N_{-}^{\downarrow} = 
	\begin{pmatrix}
	0 & 0 & 0 & 0 \\[-2pt]
	0 & 0 & 0 & 0 \\[-2pt]
	0 & 0 & 0 & 0 \\[-2pt]
	0 & 0 & 1 & 0 
	\end{pmatrix}
	,
	}
\end{equation}

\begin{equation}
	{
	\setlength\arraycolsep{3pt}
	\sigma_z =
	\begin{pmatrix}
	0 & 0 & 0 & 0 \\[-2pt]
	0 & 1 & 0 & 0 \\[-2pt]
	0 & 0 & -1 & 0 \\[-2pt]
	0 & 0 & 0 & 0 
	\end{pmatrix}
	,\quad
	\sigma_x =
	\begin{pmatrix}
	0 & 0 & 0 & 0 \\[-2pt]
	0 & 0 & 1 & 0 \\[-2pt]
	0 & 1 & 0 & 0 \\[-2pt]
	0 & 0 & 0 & 0 
	\end{pmatrix}
	,\quad
	\sigma_y =
	\begin{pmatrix}
	0 & 0 &  0 & 0 \\[-2pt]
	0 & 0 & -i & 0 \\[-2pt]
	0 & i &  0 & 0 \\[-2pt]
	0 & 0 &  0 & 0 
	\end{pmatrix}
	.
	}
\end{equation}

Using equations (\ref{eq:psi}--\ref{eq:rm1}), we can write average values 
$\left\langle A \right\rangle = \left\langle \Psi \right| A \left| \Psi \right\rangle$ for all operators in the Hamiltonian \eqref{eq:Ham0} on a given site:
\begin{eqnarray}
	\left\langle  S_z  \right\rangle &=& \cos^2 \theta \, \cos 2 \varphi , 
	\\[0.5em]
	\left\langle  S_z^2  \right\rangle &=& \cos^2 \theta , 
	\\[0.5em]
	\left\langle  S_{{+}}^2  \right\rangle &=& 
	\frac{1}{2} \, e^{-i \left(\phi_{1}-\phi_{-1}\right)} \cos ^2\theta  \sin 2\varphi , \qquad
	\left\langle  S_{{-}}^2  \right\rangle = \left\langle  S_{{+}}^2  \right\rangle ^{*} , 
	\\[0.5em]
	\big\langle  P_{{+}}^{\uparrow}  \big\rangle &=& \frac{1}{2} \, 
	e^{i \left(\phi_{\downarrow}-\phi_1\right)} \sin 2\theta  \sin \psi  \cos \varphi , \qquad 
	\big\langle P_{{-}}^{\uparrow} \big\rangle = 
	\big\langle P_{{+}}^{\uparrow} \big\rangle ^{*} , 
	\\[0.5em]
	\big\langle  P_{{+}}^{\downarrow}  \big\rangle &=& \frac{1}{2} \, 
	e^{i \left(\phi_{\uparrow}-\phi_1\right)} \sin 2\theta \cos \psi  \cos \varphi , \qquad 
	\big\langle  P_{{-}}^{\downarrow}  \big\rangle = 
	\big\langle  P_{{+}}^{\downarrow}  \big\rangle ^{*} , 
	\\[0.5em]
	\big\langle  N_{{+}}^{\uparrow}  \big\rangle &=& \frac{1}{2} \, 
	e^{-i \left(\phi_{\uparrow}-\phi_{-1}\right)} \sin 2\theta \cos \psi \sin \varphi , \qquad 
	\big\langle  N_{{-}}^{\uparrow}  \big\rangle = 
	\big\langle  N_{{+}}^{\uparrow}  \big\rangle ^{*} , 
	\\[0.5em]
	\big\langle  N_{{+}}^{\downarrow}  \big\rangle &=& \frac{1}{2} \, 
	e^{-i \left(\phi_{\downarrow}-\phi_{-1}\right)} \sin 2\theta \sin \psi \sin \varphi , \qquad 
	\big\langle  N_{{-}}^{\downarrow}  \big\rangle = 
	\big\langle  N_{{+}}^{\downarrow}  \big\rangle ^{*} , 
	\\[0.5em]
	\left\langle \sigma_x \right\rangle &=& 
	\sin^2 \theta  \sin 2\psi  \cos \left(\phi_{\downarrow}-\phi_{\uparrow} \right) , 
	\\[0.5em]
	\left\langle \sigma_y \right\rangle &=& 
	\sin^2 \theta  \sin 2\psi  \sin \left(\phi_{\downarrow}-\phi_{\uparrow} \right) , 
	\\[0.5em]
	\left\langle \sigma_z \right\rangle &=& \sin^2 \theta  \cos 2\psi . 
\end{eqnarray}

\bibliographystyle{iopart-num}
\bibliography{Manuscript_Fullmodel_CMC}

\end{document}